\documentclass[11pt,twoside]{article}
\usepackage{amsmath}
\usepackage{asp2010}
\usepackage[multiple]{footmisc}
\resetcounters

\usepackage{threeparttable}
\begin{document}

\title{AstroCloud, a Cyber-Infrastructure for Astronomy Research: Data Access and Interoperability}
\author{Dongwei Fan$^1$, Boliang He$^1$, Jian Xiao$^2$, Shanshan Li$^1$, Changhua Li$^1$, Chenzhou Cui$^1$, Ce Yu$^2$, Zhi Hong$^2$, Shucheng Yin$^2$, Chuanjun Wang$^3$, Zihuang Cao$^1$, Yufeng Fan$^3$, Linying Mi$^1$, Wanghui Wan$^{1,4}$, Jianguo Wang$^3$}
\affil{
$^1$National Astronomical Observatories, Chinese Academy of Sciences(CAS), 20A Datun Road, Beijing 100012, China\\
$^2$Tianjin University, 92 Weijin Road, Tianjin 300072, China\\
$^3$Yunnan Astronomical Observatory, CAS, P.0.Box110, Kunming 650011, China\\
$^4$Central China Normal University, 152 Luoyu Road, Wuhan 430079, China
}

\begin{abstract}
Data access and interoperability module connects the observation proposals, data, virtual machines and software. According to the unique identifier of PI (principal investigator), an email address or an internal ID, data can be collected by PI¡¯s proposals, or by the search interfaces, e.g. conesearch. Files associated with the searched results could be easily transported to cloud storages, including the storage with virtual machines, or several commercial platforms like Dropbox. Benefitted from the standards of IVOA (International Observatories Alliance), VOTable formatted searching result could be sent to kinds of VO software. Latter endeavor will try to integrate more data and connect archives and some other astronomical resources.
\end{abstract}

\section{Introduction}\label{intro}
\indent
Astronomy archives are more and more bigger. Some tasks become difficult to be done on personal computer, due to the limitation of the storage, network or the computation capability. AstroCloud of the China-VO~\footnote{China-VO~\url{http://www.china-vo.org}}\footnote{AstroCloud~\url{http://astrocloud.china-vo.org}} tries to put all the resources, including catalogs database, files, software, virtualized computers and so on to the cloud platform. Then astronomers can complete their job from the observation proposal to paper publish on the cloud, without concerning how to get the data, where to backup kinds of data, or how to configure the powerful hardware et al. The resource that astronomers need will be collected/mounted to their own virtual machine with their familiar operation system, then all the work could be done on the cloud via the remote connection.

Data access and interoperability module~\footnote{AstroCloud~Data~\url{http://explore.china-vo.org}} is a bridge to connect the resources on the cloud. A typical observation starts from the proposal, then the observed data are stored by the archiving system. Astrnomers can find the data on the data access module and transport files to their own virtual machine or just download the package to local computer. Since not all the data are published immediately, the data access module also provide an access control mechanism. Other astronomers could apply for the data and the PI of the data could check and determine whether to approve or not.

The interoperability module provides another ability that connects the web page interface with local softwares which support IVOA-SAMP protocol~\citep{taylor2012simple}, e.g. Aladin, Topcat. IVOA Simple Conesearch and Simple Spectral Access Protocol interface are also provided. Some of the interface could be found on the IVOA registers.

In section~\ref{archi}, architecture of the data access and interoperability module is provided.
This is followed by the diagram of how data flows in the Astrocloud in section~\ref{prop}.
The data access control mechanism is described in section~\ref{author}.
Conclusion is presented in section~\ref{concl}.

\section{Architecture}\label{archi}
\begin{figure}
  \centering
  \includegraphics[width=.8\textwidth]{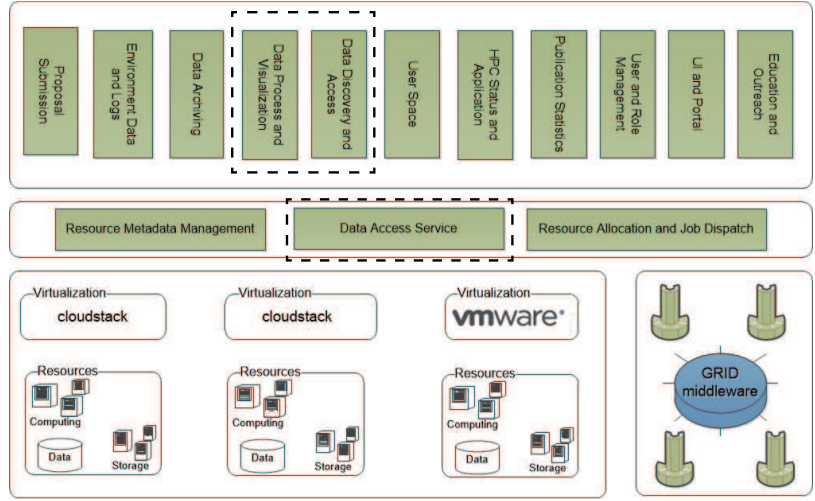}\\
  \caption{Architecture of the AstrolCloud. The general role of the data access and interoperability module are marked by the dotted rectangles.}\label{fig:archi}
\end{figure}

\indent
AstroCloud contains virtual machines, observation proposal submission, data archiving, HPC status \& applications and some other subsystems. The role of data access and interoperability module in the AstroCloud is providing data access service, discovery interface and some processes like visualization, as Fig.~\ref{fig:archi} shows. The target is to supply an easy way to help astronomers to find the data they want, including their own observation data.

\begin{figure}
  \centering
  \includegraphics[width=.6\textwidth]{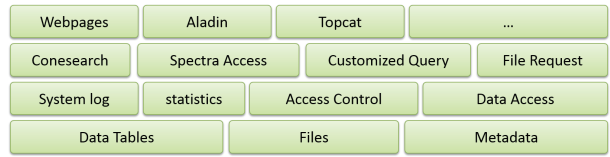}\\
  \caption{Architecture diagram of the data access and interoperability module. Several IVOA protocols are supported in this module, so users can invoke VO-driven applications to access the data besides the webpage interface.}\label{fig:archiex}
\end{figure}

In order to achieve the objective, we designed a structure in Fig.~\ref{fig:archiex}. The essential resource is the observation data and other astronomy archives. These resources should not be directly accessed by users, due to the data protection requirement. Hence data access under control is needed, and also the query logs and statistics. Several IVOA protocols are supported as well. On the webpage interface, the web-SAMP could send the searched objects list to SAMP supported clients, e.g. Aladin, Topcat. The list are also formatted in VOTable according to the Simple Conesearch or the Simple Spectra Access Protocol. This is one part of what we call ``interoperability".

\section{From Proposal to Personal Cloud Storage}\label{prop}
\begin{figure}
  \centering
  \includegraphics[width=.9\textwidth]{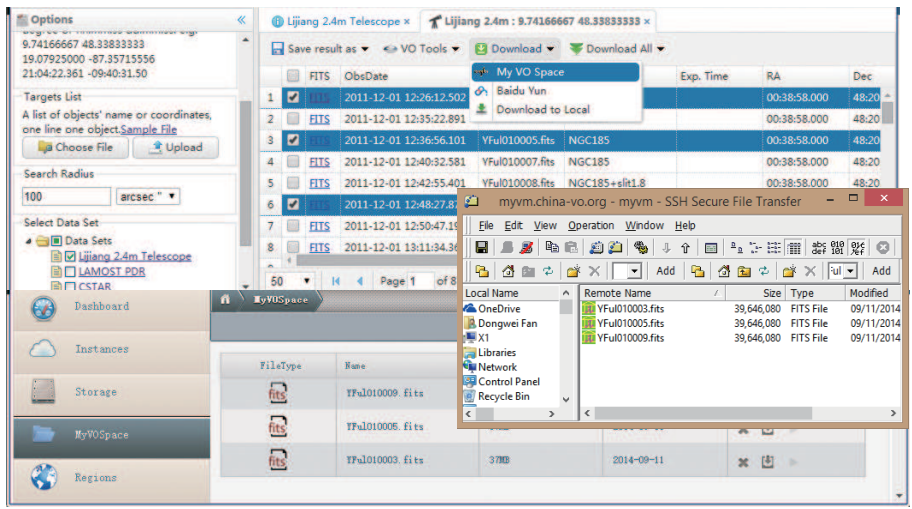}\\
  \caption{Data can be easily and very fast transported inside the cloud platform, while downloading to local disk is also supported. When choosing download data to the cloud, data will soon present on the created virtual machine or the web client. Thus astronomers could just handle the files on the virtual machine without spend lots of time waiting for the downloading or worrying about the local storage and calculation capacity.}\label{fig:query}
\end{figure}

\indent
The whole AstroCloud system maintain a user table and single sign-on mechanism, which is very beneficial to exchange information among physical individual modules. The unique ID of a user is the key to connect all the resources on the AstroCloud, from observation proposal to the archiving data, to personal private cloud storage. In the data access module, the ID is used to transport the searched files to the cloud storage. Fig.~\ref{fig:query} demonstrates that the files could be seen on the cloud storage web page, or even in the virtual machine that user created. The cloud storage directories are automatically mounted to the virtual machine. Therefore, astronomers do not have to download the huge amount of files which take very long time via internet and occupy the whole local disk. Since the storage space and the data access module are on the cloud, sometimes even in the same hardware, the files transmission speed is much more faster than normal connection. Then astronomers could analysis the data on the virtual machine. This is the way we try to solve the data flood in astronomy. ``If there is too much data to move around, take the analysis to the data."\citep{gray2003online}

\section{Data Access Control}\label{author}
\begin{figure}
  \centering
  \includegraphics[width=.7\textwidth]{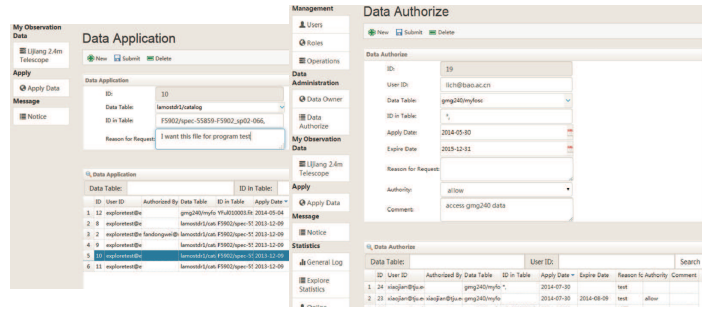}\\
  \caption{Applicant could make a request by the application page at left figure. Then data owner could check the application and grant the authority or not.}\label{fig:author}
\end{figure}

\indent
Observation data normally have a protection period, e.g. one year or 18 months. During this dates, the data only belongs to the observation proposal PI. The simple way is only allowed the PI to access the data, or even hide the data from public. But telescope managers would like to publish the data to promote the instruments' influence. One solution is to make a mechanism: if somebody interest in the data, they can write an application to access the data. Then the requirements would be delivered to the data owner, i.e. the PI. If the owner accepted the requests, the applicant could access the data. The simple process shows in Fig.~\ref{fig:author}.

\section{Conclusion}\label{concl}
\indent
Big data is not only the trend in astronomy, it is influencing the research. More and more work become difficult to be handled on personal computation devices. To work on the cloud would be a solution, and the AstroCloud of China-VO is one of the attempt. The data access and interoperability module of AstroCloud tries to collect the required data and put them to the cloud storage. To connect data from the observation proposals to astronomers' personal cloud storage, then interoperate with other modules/tools.

\acknowledgements This paper is funded by National Natural Science Foundation of China (U1231108), Ministry of Science and Technology of China (2012FY120500), Chinese Academy of Sciences (XXH12503-05-05). Data resources are supported by Chinese Astronomical Data Center.

\bibliographystyle{asp2010}	
\bibliography{P1-3}
\end{document}